\documentclass[twocolumn]{aastex61}

\usepackage{amsmath}
\usepackage{graphicx}

\usepackage{epsfig}
\usepackage{natbib}
\usepackage{wasysym}
\usepackage{lipsum}
\usepackage{mathrsfs}
\usepackage{float}
\usepackage{rotating}
\usepackage{graphicx}
\usepackage{epstopdf}
\usepackage{array}
\newcolumntype{P}[1]{>{\centering\arraybackslash}p{#1}}
\newcolumntype{M}[1]{>{\centering\arraybackslash}m{#1}}
\newcounter{chem}
\newcounter{temp}

\usepackage{amsmath}
\usepackage{mhchem}

\newenvironment{chequation}{%
  \setcounter{temp}{\value{equation}}%
  \setcounter{equation}{\value{chem}}%
  \renewcommand{\theequation}{R\arabic{equation}}%
}{%
  \setcounter{chem}{\value{equation}}%
  \setcounter{equation}{\value{temp}}%
}

\usepackage{color}

\usepackage{gensymb}

\usepackage{filecontents}

\begin{document}

\bibliographystyle{apj}

\shorttitle{Tidal-Locking \& Biosignatures}
\shortauthors{Chen, H., Wolf, E. T., et al.}

\title{Biosignature Anisotropy Modeled on Temperate Tidally Locked M-dwarf Planets}

\author[0000-0003-1995-1351]{Howard Chen}

\affil{Department of Earth and Planetary Sciences, Northwestern University, Evanston, IL 60202, USA}
\affil{Center for Interdisciplinary Exploration \& Research in Astrophysics (CIERA), Evanston, IL 60202, USA}

\author[0000-0002-7188-1648]{Eric T. Wolf}

\affil{Laboratory for Atmospheric and Space Physics, Department of Atmospheric and Oceanic Sciences, University of Colorado Boulder, Boulder, CO 80309, USA}
\affil{NASA Astrobiology Institute Virtual Planetary Laboratory, Seattle, WA 98194, USA}

\author[0000-0002-5893-2471]{Ravi Kopparapu}
\author[0000-0003-0354-9325]{Shawn Domagal-Goldman}

\affil{NASA Astrobiology Institute Virtual Planetary Laboratory, Seattle, WA 98194, USA}

\affil{NASA Goddard Spaceflight Center, Greenbelt, MD 20771, USA}
\author[0000-0002-2065-4517]{Daniel E. Horton}

\affil{Department of Earth and Planetary Sciences, Northwestern University, Evanston, IL 60202, USA}
\affil{Center for Interdisciplinary Exploration \& Research in Astrophysics (CIERA), Evanston, IL 60202, USA}

\begin{abstract}
A planet's atmospheric constituents (e.g., O$_2$, O$_3$, H$_2$O$_v$, CO$_2$, CH$_4$, and  N$_2$O) can provide clues to its surface habitability, and may offer biosignature targets for remote life detection efforts. The plethora of rocky exoplanets found by recent transit surveys (e.g., the {\it Kepler} mission) indicates that potentially habitable systems orbiting K- and M-dwarf stars may have very different orbital and atmospheric characteristics than Earth. To assess the physical distribution and observational prospects of various biosignatures and habitability indicators, it is important to understand how they may change under different astrophysical and geophysical configurations, and to simulate these changes with models that include feedbacks between different subsystems of a planet's climate. Here we use a three-dimensional (3D) Chemistry-Climate model (CCM) to study the effects of changes in stellar spectral energy distribution (SED), stellar activity, and planetary rotation on Earth-analogs and tidally-locked planets. Our simulations show that, apart from shifts in stellar SEDs and UV radiation, changes in illumination geometry and rotation-induced circulation can influence the global distribution of atmospheric biosignatures. We find that the stratospheric day-to-nightside mixing ratio differences on tidally-locked planets remain low ($<20\%$) across the majority of the canonical biosignatures. Interestingly however, secondary photosynthetic biosignatures (e.g., C$_2$H$_6$S) show much greater (${\sim}67\%$) day-to-nightside differences, and point to regimes in which tidal-locking could have observationally distinguishable effects on phase curve, transit, and secondary eclipse measurements. Overall, this work highlights the potential and promise for 3D CCMs to study the atmospheric properties and habitability of terrestrial worlds.
\end{abstract}

\keywords{astrobiology -- planets and satellites: atmospheres --  planets and satellites: terrestrial planets}

\correspondingauthor{Howard Chen, Northwestern University}
\email{howard@earth.northwestern.edu}

\section{Introduction}

A promising approach in the hunt for life beyond Earth is through the detection of biosignatures -- biologically produced compounds such as O$_2$, CH$_4$, N$_2$O, and CO$_2$ -- in the atmospheres of terrestrial planets orbiting the putative habitable zones (HZs) of nearby stars \citep{Lovelock1975,SaganEt1993NATURE,SeagerEt2009,KastingEt2015ApJL}.

In recent years, the convergence of our ability to detect, confirm, and characterize extrasolar planets has profoundly strengthened the prospects of finding life on other worlds. Consistently improving measurements of stellar mass, radius, and distance allows more accurate constraints on their attending planets \citep{MannEt2015ApJ}. Large-scale observational surveys such as the M-Earth project, TRAPPIST survey, Hungarian Automated Telescope Network (HATNet), Kepler Space Telescope, and Transiting Exoplanet Satellite Survey (TESS) have detected planets in the habitable zones around these stars \citep{ThompsonEt2018ApJS} and will continue to monitor closer and brighter systems for Earth-sized planets \citep{BarclayEt2018arXiv}. Simultaneously, follow-up characterization efforts of these confirmed planets were able to resolve atmospheres of much smaller planets than past efforts (e.g., HAT-P-26b; \citealt{WakefordEt2017SCI}). Looking ahead, a variety of instruments are being designed with life detection goals in mind. This includes ground-based observatories such as the European Extremely Large Telescope (E-ELT), Giant Magellen Telescope (GMT), and Thirty-Meter Telescope (TMT), as well as space-based missions such as the James Webb Space Telescope (JWST), Large UV/Optical/IR Surveyor (LUVOIR), Origins Space Telescope (OST), and Habitable Exoplanet Imaging Observatory (HabEx). HabEx and LUVOIR in particular would enable characterization of potentially habitable Earth-sized rocky planets in our solar neighborhood ($\la 100$ parsecs; \citealt{BolcarEt2016PROC,MennessonEt2016PASP,BatalhaEt2018ApJL}). 

The recent discoveries of Proxima Centauri b \citep{anglada2016terrestrial} and the TRAPPIST-1 system \citep{GillonEt2017NATURE} demonstrate that analyses of small rocky planets are within reach. However, many of these planets orbit extremely close to their host M-type stars (0.02-0.2 AU) and are susceptible to trapping by tidal-forces \citep{TarterEt2007}. Tidally-locked but potentially habitable planets are expected to be common in HZs of low-mass stars (${\sim} 15\%$; \citealt{Dressing+Charbonneau2015ApJ}) -- which dominate our solar neighborhood stellar population (${\sim} 70\%$; \citealt{HenryEt2006AJ}). Concurrently, our earliest opportunity for a biosignature search will likely come from the JWST and ground-based extremely large telescopes (E-ELT, GMT, and TMT); these observatories will enable spectroscopic observations of rocky planets around K- and M-type stars. It is therefore likely that our first opportunity to measure atmospheres of rocky worlds will be tidally-locked terrestrial planets around K- or M-dwarf stars.

Characterization of exoplanets primarily involves measuring starlight and terrestrial thermal emissions absorbed by planetary atmospheres as a function of wavelength. For transit spectroscopy, which will be the main tool for obtaining spectra from planets around M-dwarf stars, observations are biased towards atmospheric constituents across the terminators. Therefore, interpreting spectroscopic observations requires inferring both the concentration and distribution of detectable gases. Such properties can be predicted by 3D global climate and chemistry-climate models (GCMs and CCMs). GCMs and CCMs are numerical models that employ laws of physics, fluid motion, and in the case of CCMs, chemistry to simulate movements, interactions, and climatic implications of a planet's atmospheric constituents and boundary conditions.

Previous simulations of atmospheres of tidally-locked planets performed with 3D GCMs have demonstrated that habitable states of tidally-locked planets are strong functions of: ({\it i}) Coriolis force \citep{YangEt2017ApJ,WayEt2016GRL,KopparapuEt2016ApJ},  ({\it ii}) stellar energy distribution (SED) and bolometric stellar flux \citep{KopparapuEt2017ApJ,Wolf2017ApJL}, ({\it iii}) atmospheric mass \citep{Wordsworth2015ApJ}, and ({\it iv}) radiative transfer scheme \citep{YangEt2017ApJ}. Despite the ability of GCMs to simulate key climatological factors, as demonstrated by these studies, their foci have primarily been on questions of habitability, rather than the concentrations and distributions of biologically-produced gases and habitability indicators.

To study effects of tidal-locking on atmospheric chemistry and molecular spectroscopic signals, models capable of resolving chemical speciation, reactions, and transport are needed. To date, exoplanet atmospheric photochemical predictions have largely relied on one-dimensional global-mean photochemistry-climate models (e.g., \citealt{KastingEt1984,SeguraEt2005AsBio,MeadowsEt2018AsBiob}). These 1D models have been used to simulate synthetic spectra of hypothetical rocky planets under the influence of different host SEDs \citep{RauerEt2011A&A,RugheimerEt2015ApJ}. However, 1D models employ relatively simple eddy-diffusion parameterizations for vertical transport and do not account for atmospheric dynamics, climate heterogeneities, or 3D geometric effects critical to observations. These factors are important as advection and diffusion can affect concentration, distribution, and ultimately the composition of an atmosphere \citep{seinfeld2012atmospheric}. In addition to altering photochemistry, as shown by 1D models, shifts in stellar SED can influence atmospheric circulation and climate (e.g., \citealt{ShieldsEt2014ApJ,FujiiEt2017ApJ}). Atmospheric chemistry and dynamics are thus interactive, and should ideally be simulated using fully-coupled 3D model components. 

Here, to better understand the observational potential of tidally-locked planets, the integrated effects of atmospheric chemistry, photochemistry, and circulation are considered over the 3D geometry of a planet's atmosphere. In this Letter, we simulate Earth-analogs and tidally-locked planets around M-dwarf stars using a 3D CCM, while seeking to ({\it i}) elucidate the photochemical nature of Earth-like worlds, ({\it ii}) demonstrate the utility of 3D CCMs in terrestrial exoplanet studies, ({\it iii}) and advance model comparison efforts between 3D and 1D research communities.

\begin{table*}[t]   
\centering
\caption{\normalsize{{\bf Model Comparisons of Approximate Global-mean Mixing Ratios of Various Gases}}}  
\vspace{-1ex}
\begin{tabular}{|l|M{2.7cm}|M{2.7cm}|M{2.7cm}|}
\hline
Study   & Rauer et al. (2011)  & Rugheimer et al. (2015)  & This work\\ \hline
Model &  1D photochemical & EXO-P   & NCAR's CAM-chem\\ \hline
SED data       &    AD Leonis$^a$   & active M6 stellar model$^b$  & Proxima Cen.$^c$    \\ \hline
Bond Albedo &     N/A$^d$                     &  0.07      & 0.46    \\ \hline
$T_{\rm surf}$ (K)      &       $ 298$                 & $ 300$           & $242$     \\ \hline
O$_{3, {\rm surf}}$ (mol mol$^{-1}$)          &    $8 \times 10^{-11}$            &      $10^{-9}$ & $9.4 \times 10^{-13}$       \\ \hline
CH$_{4, {\rm surf}}$  (mol mol$^{-1}$)      &    $10^{-4}$               & $1 \times 10^{-3}$ & $3.4 \times 10^{-4}$      \\ \hline
N$_2$O$_{\rm surf}$  (mol mol$^{-1}$)   &       $2.0 \times 10^{-6}$                 &    $1.7 \times 10^{-6}$     & $2.4 \times 10^{-6}$          \\ \hline
H$_2$O$_{\rm surf}$  (mol mol$^{-1}$)        &      $ 7.0 \times 10^{-2}$                  &  $5.0 \time 10^{-2}$      & $3.5 \times 10^{-4}$     \\ \hline
$T_{\rm 100 mb}^e$  (K)    & $251$              & $245$      & $200$      \\ \hline
O$_{3, {\rm 100 mb}}$ (mol mol$^{-1}$)         &     $10^{-7}$                & $10^{-6}$ & $5.2 \times 10^{-7}$       \\ \hline
CH$_{4, {\rm 100 mb}}$   (mol mol$^{-1}$)         &     $10^{-5}$           &    $1 \times 10^{-3}$    & $3.1 \times 10^{-4}$      \\ \hline
N$_2$O$_{\rm 100 mb}$   (mol mol$^{-1}$)        &      $10^{-5}$                   & $5 \times 10^{-7}$           & $2.0  \times 10^{-6}$          \\ \hline
H$_2$O$_{\rm 100 mb}$  (mol mol$^{-1}$)             &     $ 10^{-6}$                       &  $7.0 \times 10^{-4}$       & $5.8\times 10^{-6}$     \\ \hline
\end{tabular}\par
\begin{flushleft} 
Approximate atmospheric temperature (units of K) and mixing ratios of various gas phase species (units of mol mol$^{-1}$) simulated on Earth-like planets by various authors.  All data reported are taken from simulations forced by active stellar SEDs that range from M6V to M8V dwarf stars. $^a $Active M3.5eV star, $T_{\rm eff} = 3300$ K, $M = 0.41 M_\odot$. $^b$Part of a suite of stellar SED model data generated by Rugheimer et al. (2015) to represent the most active M-dwarf observations (e.g., PHOENIX or MUSCLES database). $^c$SED data from Virtual Planetary Laboratory's stellar spectrum database (Meadows et al. 2018a). $^d$Data not provided. $^e$Values reported at 100 mb. 
\end{flushleft}
\end{table*}
\section{Model Description \& Experimental Setup }

In this study, we employ the Community Atmosphere Model with Chemistry (CAM-chem), a subset of the National Center for Atmospheric Research (NCAR) Community Earth System Model (CESM v.1.2), to investigate atmospheres of Earth-like planets. CAM-chem is a 3D global CCM that simulates interactions of atmospheric chemistry, radiation, thermodynamics, and dynamics (for a complete model description, see \citet{LamarqueEt2012}). CAM-chem combines the CAM4 atmospheric component with the fully implemented Model for Ozone and Related Chemical Tracers (MOZART) chemical transport model. CAM-chem resolves 97 gas phase species and aerosols linked by 196 chemical and photolytic reactions. CAM, the atmosphere component of the model, has seen wide applications in problems of paleoclimate and exoplanets  (e.g., \citealt{Wolf+Toon2015JGR,KopparapuEt2016ApJ}), whereas CAM-chem has largely been limited to studies of present Earth. All simulations presented were run for 30 Earth years and reported results are averaged over the last 20. 

We simulate Earth-analogs and tidally-locked planets and assess the sensitivity of atmospheric biosignatures to three primary variables: ({\it i}) stellar spectral energy distribution, ({\it ii}) stellar UV radiation, and ({\it iii}) planetary rotation period. To simulate Earth-analogs, we use a preindustrial Earth setup forced by solar spectral irradiance data \citep{lean1995reconstruction}, i.e., apart from the orbital parameters described below, our Earth-analog simulation uses identical boundary and initial conditions to Earth in 1850, prior to anthropogenic influences \citep{taylor2012overview}. These conditions include atmospheric gases N$_2$ (78\% by volume), O$_2$ (21\%), and CO$_2$ ($2.85 \times 10^{-2} \%$) \citep{macfarling2006law}. In addition, the model simulates the free-running evolution of H$_2$O$_v$ and O$_3$, while CH$_4$ and N$_2$O surface fluxes are latitudinally variable (global mean CH$_4$: $7.23 \times 10^{-7}$ and N$_2$O: $2.73 \times 10^{-7}$ mol mol$^{-1}$). Throughout the remainder of the paper, we refer to this Earth-Sun simulation as the baseline. 

\begin{figure*}[t]   
\begin{center}
\includegraphics[width=2.05\columnwidth]{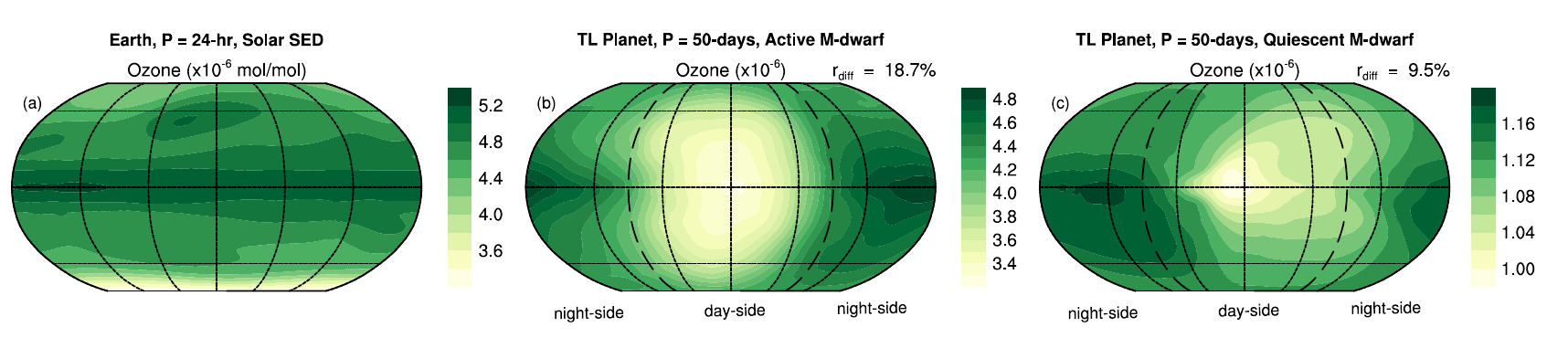}
\includegraphics[width=2.05\columnwidth]{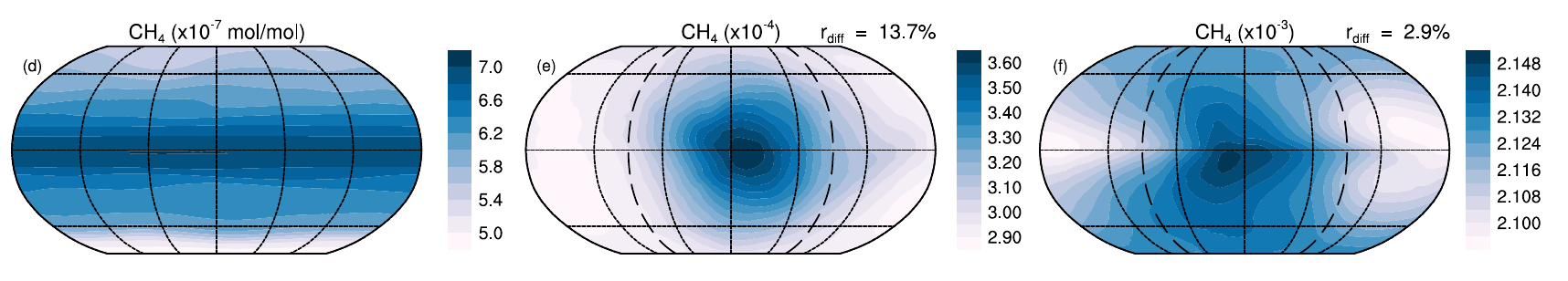}
\includegraphics[width=2.05\columnwidth]{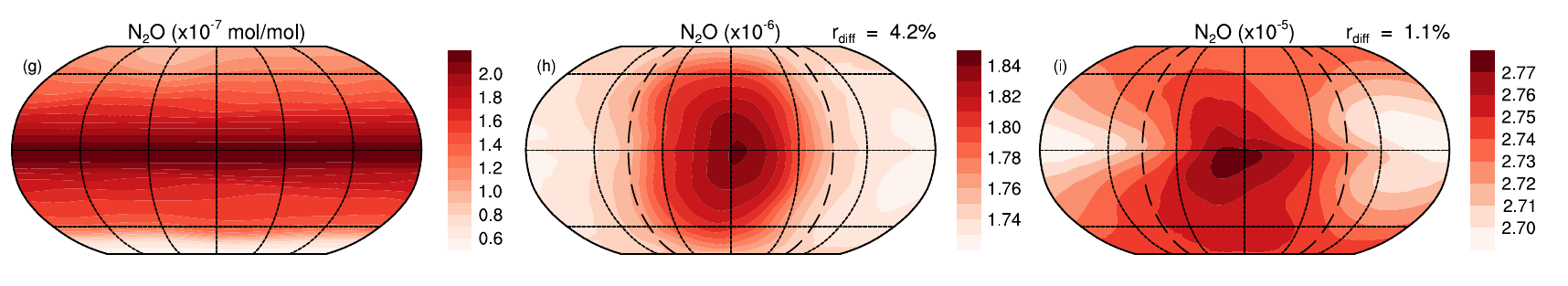}
\includegraphics[width=2.05\columnwidth]{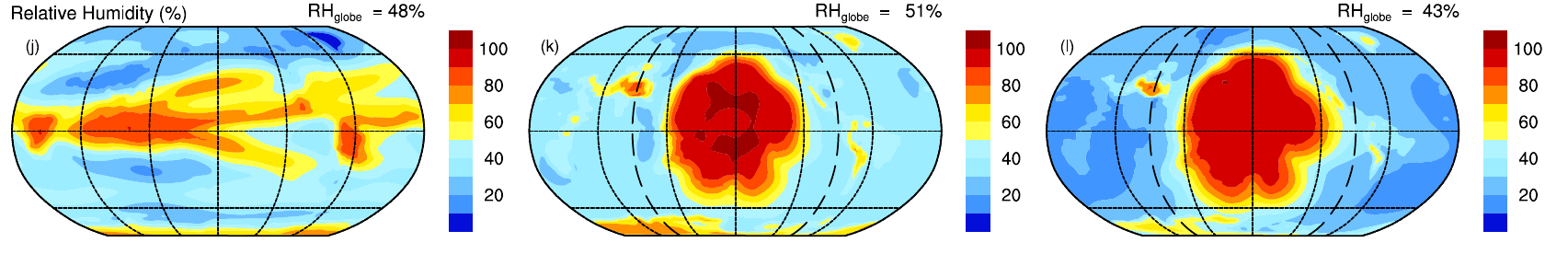}
\caption{Global distribution of O$_3$, CH$_4$, and N$_2$O mixing ratios and relative humidity for Earth-like non-tidally-locked ($P = 24$ hrs) Solar-SED simulations (a, d, g, j) and tidally-locked ($P = 50$ days) active (b, e, h, k) and quiescent (c, f, i, l) M-dwarf SED simulations. Evidence of circulation- and photochemical-induced biosignature anisotropy are apparent. Day-to-nightside mixing ratio contrasts ($r_{\rm diff}$) for tidally-locked simulations are reported, while relative humidity is averaged across the globe (RH$_{\rm globe}$). Gas mixing ratios are pressure-weighted vertical averages over the top of the model atmosphere (1-to-100 mb). Relative humidity is reported for the 200 mb pressure surface.  Note differences in scaling factors used amongst experiments and constituents. Dashed-lines indicate locations of terminators. }
\end{center}
\end{figure*}

We also modify CAM-chem to simulate tidally-locked planets with initially Earth-like atmospheric compositions forced by M-dwarf SEDs. This SED was obtained from an open-source dataset of an M6V star, Proxima Centauri, compiled by NASA's Virtual Planetary Laboratory (VPL) team and is available at \url{http://vpl.astro.washington.edu/spectra/stellar/}. We explore two SED-types (active and quiescent) that bracket the endmember ranges of stellar activity. VPL Proxima Cen. data is assumed to be moderately-to-highly active. To construct a quiescent M-dwarf SED, we swap out UV bands ($\lambda<500$ nm) of the original Proxima Cen. data with that of a low-activity star (HD114710). For all exo-Earth simulations, we assume tidal-locking (i.e., trapped in 1:1 spin-orbit resonances), with orbital periods of 50 Earth days. While we do not use self-consistent stellar-flux orbital period relationships (e.g., \citealt{KopparapuEt2016ApJ,Haqq-MisraEt2018ApJ}), the idealized case studied here highlights the value of using CCMs for modeling chemical processes on slowly and synchronously rotating planets.

For both Earth and tidally-locked exoplanet simulations, we set orbital parameters (obliquity, eccentricity, and precession) to zero, such that top-of-atmosphere incident stellar flux is symmetric about the equator. Incident bolometric stellar flux for all simulations is set to 1360 W m$^{-2}$. The substellar point for all simulations is fixed at (Earth’s) latitude = 0$\degree$ and longitude = 180$\degree$, in the Pacific Ocean. Note that other studies (e.g., \citealt{LewisEt2018ApJ}) have shown that surface type beneath a substellar point can modify water vapor availability, influencing water vapor-induced greenhouse and cloud radiative effects, and possibly atmospheric chemistry.

In all simulations, we assume present Earth’s continental configuration, topography, mass, and radius. We use prognostic atmospheric and oceanic components of CESM, as well as prescribed preindustrial land, surface ice, and sea ice components. Horizontal resolution (latitude $\times$ longitude) is set to $1.9\degree \times 2.5\degree$ with 26 vertical atmospheric levels and model top of 1 mb (${\sim}50$ km). The land model is Community Land Model version 4.0 with non-interactive surface features. The ocean component is a thermodynamic slab model with prescribed heat flux values sourced from dynamical ocean simulations (e.g., \citealt{danabasoglu2009equilibrium}). 

Consistent with 1D studies (e.g., Segura et al. 2005, Rugheimer et al. 2015) and in alignment with our lack of terrestrial exoplanet observations, we assume atmospheric compositions, biological production, and dry deposition rates of gaseous species the same as those of preindustrial Earth. Apart from CH$_4$ and N$_2$O, global surface gas flux inputs are based on spatially-explicit preindustrial monthly averages (e.g., DMS; \citealt{kettle2000flux}). Due to SED sensitivities, CH$_4$ and N$_2$O surface flux boundary conditions are estimated via ancillary CCM simulations that allow for the emergence of stellar SED-dependent flux magnitudes (i.e., WACCM; \citealt{neale2010description}). Emergent SED-consistent N$_2$O and CH$_4$ flux estimates are temporally and spatially fixed in active and quiescent M-dwarf simulations at CH$_4$: $3.5 \times 10^{-4}$ and $2.3 \times 10^{-3}$ mol mol$^{-1}$ and N$_2$O: $2.5\ \times 10{^-6}$ and $3.2 \times 10^{-5}$ mol mol$^{-1}$, respectively. Given uncertainties inherent in flux estimates, sensitivity experiment and day-to-nightside mixing ratio comparisons should focus on relative rather than absolute differences.

\section{Results}

Three general observations can be made from our simulated 3D global distributions of O$_3$, CH$_4$, N$_2$O, and DMS on Earth-like and tidally-locked planets (Figures 1 and 4): ({\it i}) Changes in mixing ratios of O$_3$, CH$_4$, and N$_2$O are primarily due to different levels of stellar UV flux amongst the three SED datasets. ({\it ii}) Introduction of tidal-locking modifies globally homogeneous gas distributions that characterize Earth-like scenarios. ({\it iii}) Heterogeneous surface-to-atmosphere flux distributions (e.g., DMS) can influence the resultant mixing ratios of atmospheric constituents. 

To facilitate analysis of our results, we define a day-to-nightside mixing (mole) ratio contrast as:
\begin{equation}
    r_{\rm diff}  = \frac{r_{\rm day} - r_{\rm night}}{r_{\rm globe} }
\end{equation}

i.e., the relative difference between the two hemispheres, where $r_{\rm day}$ is the dayside hemispheric mixing ratio mean, $r_{\rm night}$ the nightside mean, and $r_{\rm globe}$ the global mean. The degree of anisotropy is loosely encapsulated in this parameter, which is analogous to the definition used by \citet{Koll+Abbot2016ApJ} in the context of temperature contrasts. Values of $r_{\rm diff}$ for each respective experiment are shown in Figures 1 and 4 and will be discussed throughout the paper.

\begin{figure*}[t]   
\begin{center}
\includegraphics[width=2\columnwidth]{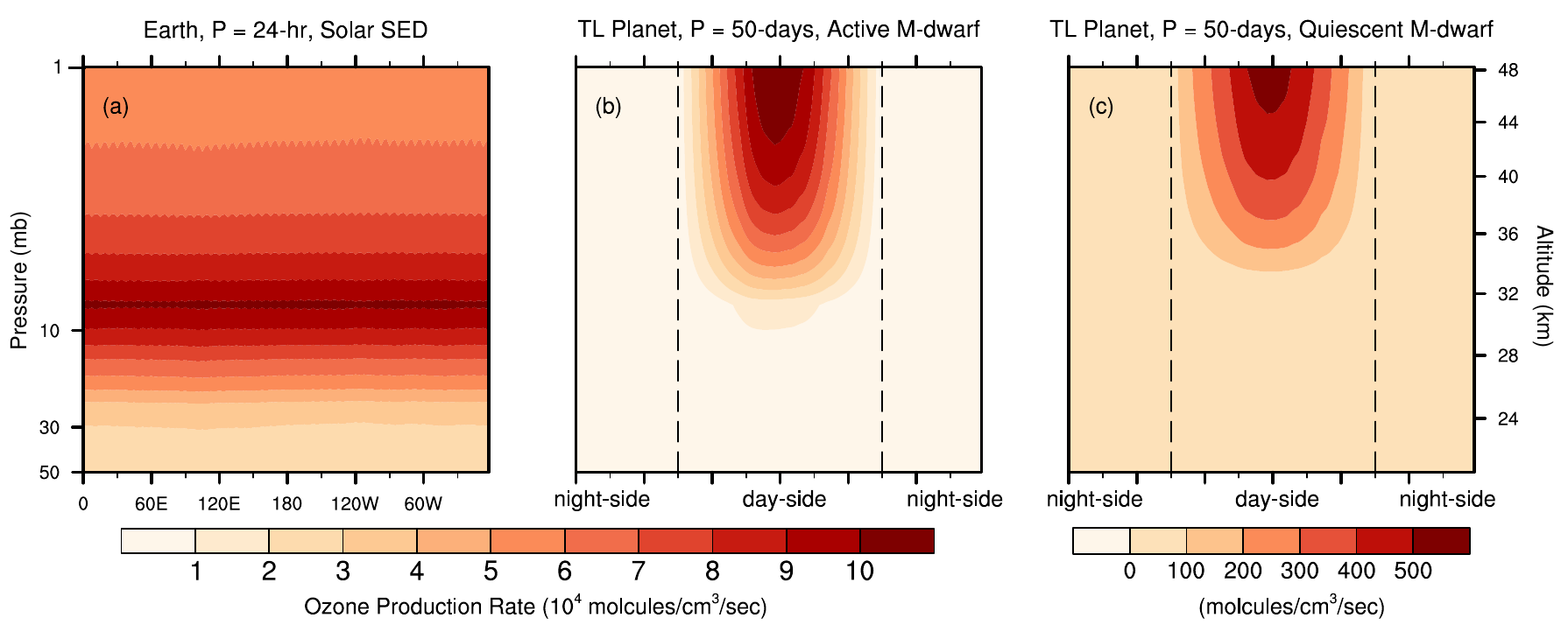}
\includegraphics[width=2\columnwidth]{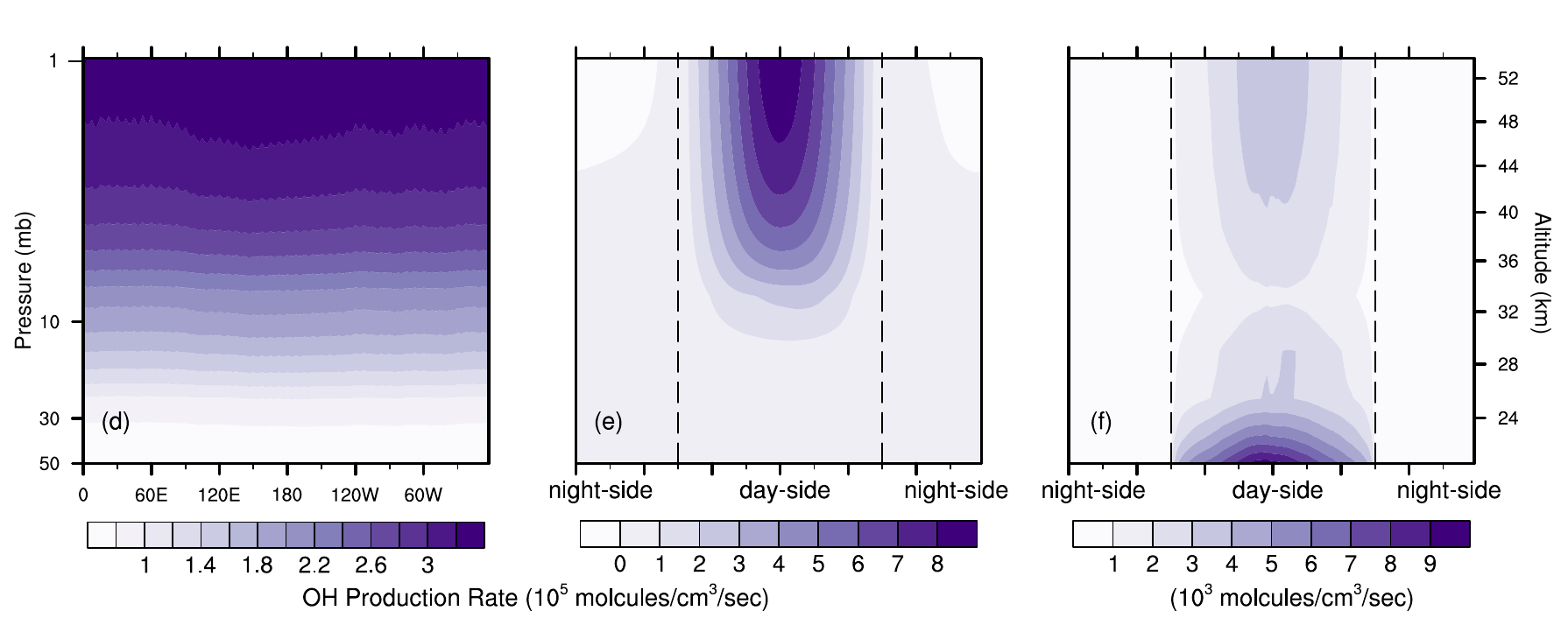}
\caption{Stratospheric O$_3$ and OH production rates as functions of longitude for Earth-like non-tidally-locked ($P = 24$ hrs) Solar-SED simulations (a, d) and tidally-locked ($P = 50 $ days) M-dwarf SED (b, c, e, f) simulations. Photolytic processes drive ozone and hydroxyl radical production and help to explain many of the observed biosignature gas distributions in Figure 1. Note vertical axis begins at 50 mb (${\sim}20$ km).  Dashed-lines indicate locations of terminators.}
\end{center}
\end{figure*}

\vspace{6ex}

\subsection{Ozone Distributions, Water Vapor Mixing Ratios, and Temperature Profiles}

Ozone production and destruction depend on stellar UV activity, availability of molecular and atomic oxygen, and ambient meteorological conditions $(P, T)$. As our simulated M-dwarf SED is moderately active in the UV bands, our results show similar quantities of ozone between the baseline Earth-Sun and tidally-locked cases (Figures 1a-c). However, the quiescent SED, produces lower ozone concentrations above the tropopause (Figure 3c). 
These differences reflect specific stellar activity inputs. Quiescent M-dwarfs emit lower UV in the range responsible for ozone production ($160< \lambda < 240$ nm). Moreover, calculated day-to-nightside mixing ratio differences $r_{\rm diff}$ are higher (${\sim} 19\%$) in the active M-dwarf SED scenario.

Modulations to ozone concentration have major influences on distributions of other biosignature gases. This is due to substellar hemispheric production of excited state atomic oxygen O($^1$D) and constituent families of HO$_x$. Both O($^1$D) and HO$_x$ constituents are reactive radicals important for atmospheric biogenic organo-compounds and hydrocarbons (e.g., CH$_4$, CH$_3$, HCL, H$_2$S). 

As ozone is photochemically produced, horizontal advection carries a portion to the nightside as evidenced by its presence in both hemispheres. The lifetime of ozone (${\sim} 15$ days), in conjunction with day-to-nightside transport, is sufficient enough to sustain some nightside O$_3$, but not efficient enough to fully mix the atmosphere, allowing day-to-nightside contrast (${\sim} 19\%$; Figure 1b). Conversely, the product O($^1$D) shows limited transport effects due to a short lifetime ($<5$ secs), reflected in its large $r_{\rm diff}$ value (${\sim} 300\%$; not shown). O($^1$D) is rapidly removed by either the R1 or R2 reaction pathway \citep{jacob1999introduction}:
 
\begin{chequation}
\begin{align}
{\rm O(}^1{\rm D)} + {\rm H}_2{\rm O}_v \rightarrow 2{\rm OH}                 \\
{\rm O(}^1{\rm D)} + {\rm M(N}_2{\rm , O}_2{\rm )} \rightarrow {\rm O(}^3{\rm P)} + {\rm M}                  \\
{\rm O(}^3{\rm P)} + {\rm O}_2 + {\rm M} \rightarrow  {\rm O}_3 + {\rm M}                 
\end{align}
\end{chequation}

The higher ozone mixing ratios on tidally-locked nightsides is explained by these reaction pathways (Figure 1b-c). Reaction R1, which creates the hydroxyl radical OH, predominantly occurs on the dayside due to the abundance of H$_2$O$_v$ (Figures 2e-f). In R2, singlet oxygen returns to the triplet ground state, which can then recombine with oxygen to form ozone via R3. Considered together, significant dayside UV ozone destruction and enhanced removal of O($^1$D) by water vapor offset higher ozone production rates (Figure 3b), which helps to explain lower dayside ozone mixing ratios (Figure 1b-c).

Interaction of stellar UV photons with O$_2$ and O$_3$ can also be seen in vertical temperature profiles. On Earth-like planets, stratospheric temperature is primarily a function of incident UV flux ($200< \lambda < 310$ nm) due to the role of O$_3$ absorption of shortwave photons. In our simulations, this feature is apparent in global and hemispherically averaged profiles. Our simulations indicate that upper-stratospheric temperatures increase (and inversions weaken) as UV radiation levels increase; from quiescent SED, to active SED, to the baseline Earth simulation (Figure 3a). Enhanced UV absorption by O$_3$ and O$_2$ increase temperatures above the tropopause, reducing the vertical gradient and inversion strength. 

We now turn to discussing testability of our 3D model predictions. Based on simulated ozone distributions, the calculated $r_{\rm diff}$ (${\sim} 20\%$) is notable but unlikely to be discernible with current observational capabilities \citep{Burrows2014PNAS}. However, this task may prove viable using future instruments (e.g., \citealt{GreeneEt2016ApJ}). \citet{ProedrouEt2016} reached a similar conclusion by comparing total column ozone of a tidally-locked Solar-SED Earth and found a ${\sim} 23\%$ difference between mean ozone columns during four arbitrary phases due to varying viewing angles. 

Compared with 1D model studies, our 3D simulations produce similar ozone mixing ratios (Table 1). However, we find substantially different Bond albedos, temperature, and water mixing ratio profiles. As a consequence of increased dayside water vapor-induced opacity and substellar clouds on tidally-locked planets, global-mean surface temperatures of both tidally-locked simulations are ${\sim}40$ K colder than the baseline (Figure 3a), while Bond albedos are substantially higher (Table 1), in agreement with GCM studies \citep{YangEt2013ApJL,KopparapuEt2016ApJ,KopparapuEt2017ApJ}. Colder global (and nightside) temperatures produce lower global water vapor mixing ratios than predicted by 1D models with clear-sky assumptions (i.e., pure water vapor without clouds). Conversely, dayside H$_2$O$_v$ mixing ratios are greater due to humid updrafts at the substellar point (Figures 1j-l). Curiously, the quiescent M-dwarf SED simulation (Figure 1l) has lower global-mean relative humidity than the Earth-analog (Figure 1j) and the active M-dwarf case (Figure 1k). This is due to increased ozone mixing ratios and degree of UV absorption, which limit the photolysis of H$_2$O$_v$, in the more active SED simulations. Hence counterintuitively, more dayside H$_2$O$_v$ destruction is experienced by the simulation under lower UV radiation. Such behaviors exemplify the value of CCM simulations, in which capturing feedbacks between 3D dynamical processes, solar forcing, and atmospheric chemistry is critical.

\begin{figure*}[t]   
\begin{center}
\includegraphics[width=1.95\columnwidth]{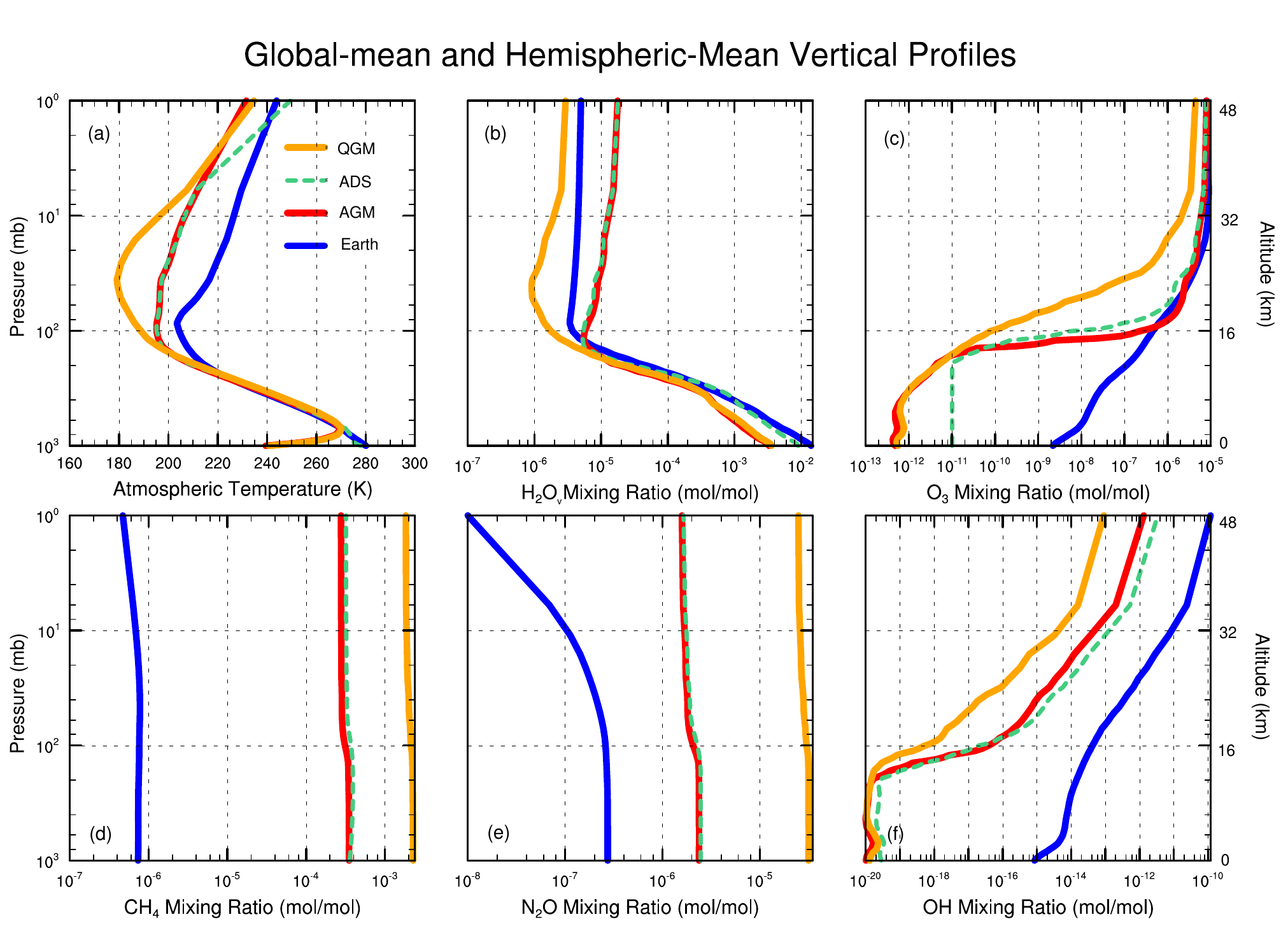}
\caption{ Vertical profiles of global-mean temperature (a) and mixing ratios of various gas phase species (b, c, d, e, f). QGM (quiescent global-mean) denotes global-mean values from simulations forced by a quiescent, AGM (active global-mean) denotes those forced by an active M-dwarf SED, and ADS (active day-side) denotes dayside-mean values from simulations forced by an active M-dwarf SED. Note axes are log-scaled and begin at planetary surfaces (${\sim} 1000$ mb).}
\end{center}
\end{figure*}

\subsection{Types I and II Biosignatures: 3D CH$_4$ and N$_2$O Abundances}

CH$_4$ and N$_2$O are important biosignatures produced by a myriad of bacterial metabolic pathways \citep{DesMaraisEt2002AsBio,SchweitermanEt2018AsBio}. In Figures 1d-f, we show modeled CH$_4$ distributions. High CH$_4$ mixing ratios for planets orbiting quiet M-dwarfs were first noted by \citet{SeguraEt2005AsBio} using a 1D model, caused by reduced photochemical removal by reaction with OH:

\vspace{-1ex}
\begin{chequation}
\begin{align}
{\rm CH}_4 + {\rm OH} \rightarrow  {\rm CH}_3 + {\rm H}_2{\rm O}     
\end{align}
\end{chequation}
\vspace{-1ex}
 
In our 3D CCM, we find similar global mean CH$_4$ increases in tidally-locked simulations (Figures 1e-f). However, active and quiescent simulations have low CH$_4$ $r_{\rm diff}$ values (13.7\% and 2.9\%, respectively). Low $r_{\rm diff}$ values are explained by a mixture of competing processes. First, upwelling in tidally-locked simulations occurs exclusively below the substellar point, as evidenced by upper-tropospheric moisture patterns (Figures 1j-l). Compared to the baseline, tidally-locked meridional overturning circulation is strengthened, which brings greater moisture aloft. This, in conjunction with the dayside abundance of OH, removes CH$_4$ via R4. Increased dayside OH production (Figures 2e-f) is a consequence of abundant O($^1$D) and H$_2$O$_v$ (R1), both of which are sparse on the nightside. These processes combine to limit dayside CH$_4$ and produce lower $r_{\rm diff}$ values than expected.

N$_2$O is primarily destroyed by UV photons ($\lambda<220$ nm) and photo-oxidation by reactions with stratospheric O($^1$D) (Figure 1g-i). Hence predicted N$_2$O concentrations around active M-dwarfs are lesser than those with quiescent SEDs. For both active and quiescent SED simulations, higher concentrations within the substellar hemisphere are found (Figures 1h-i), similar to CH$_4$ behavior. 

Interestingly, simulations forced by the active SED have greater stratospheric $r_{\rm diff}$ (CH$_4$: 13.7\% and N$_2$O: 6.9\%; Figures 1e-h) than those forced by quiescent SEDs (CH$_4$: 2.9\% and N$_2$O: 1.1\%; Figures 1f-i). Higher $r_{\rm diff}$ values for active SED cases is somewhat counterintuitive as one might expect that enhanced photolytic destruction on planets around active M-dwarfs should suppress day-to-nightside contrasts. However, simulations forced by active SEDs have more isothermal atmospheres (i.e., weaker temperature inversions; Figure 3a), which promote vertical mixing of surface gases above the tropopause, contributing to higher $r_{\rm diff}$ values. 
\begin{figure*}[t]   
\begin{center}
\includegraphics[width=2.0\columnwidth]{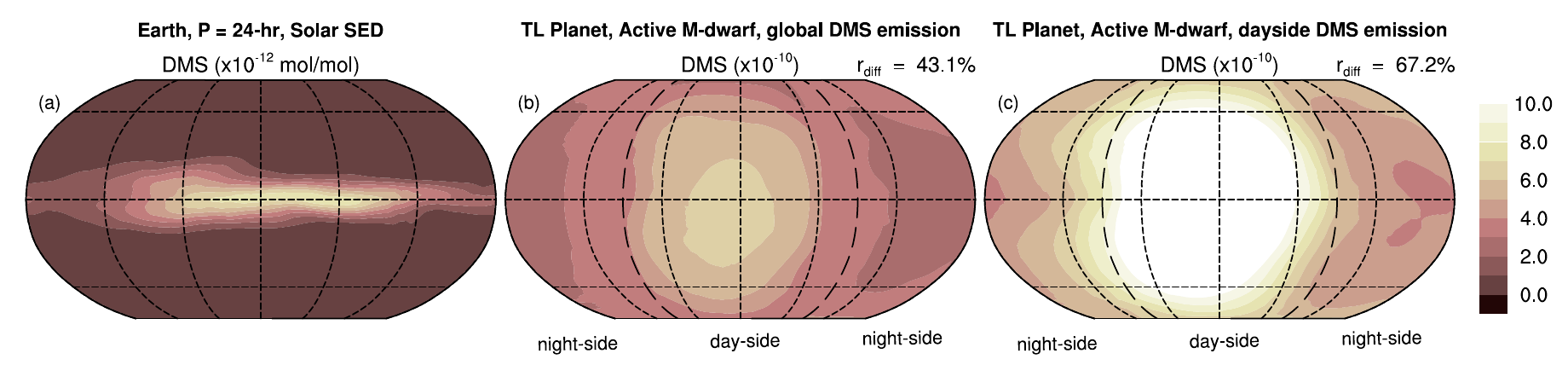}
\caption{Global distribution of photosynthetic biosignature dimethyl sulfide (DMS, (CH$_3$)$_2$S) for Earth-like non-tidally-locked ($P = 24$ hrs) Solar-SED simulations (a), tidally-locked ($P = 50$ days) active M-dwarf SED simulations with a global DMS flux assumption (b), and tidally-locked with a dayside DMS flux assumption (c). Day-to-nightside mixing ratio differences ($r_{\rm diff}$) for the tidally-locked simulations are reported. Phototrophs are assumed to be only present on the permanently lit day-side in panel (c), which results in an enhanced stratospheric day-to-nightside mixing ratio contrast. Dashed-lines indicate locations of terminators.}
\end{center}
\end{figure*}

\subsection{Effects of Surface Fluxes on Atmospheric Distribution: Case of Dimethyl Sulfide}

On tidally-locked planets, phototrophs are unlikely to emit biogenic gases globally (i.e., the assumption for all biosignature gases considered thus far); rather, 
ally-derived emissions are likely to be restricted to the dayside. To see how a biosignature gas (e.g., DMS; C$_2$H$_6$S) may behave on a tidally-locked planet, we conduct three experiments, each with a different DMS flux distribution assumption, i.e., Earth-like, tidally-locked with global DMS flux, and tidally-locked with dayside DMS flux (Figure 4). We find that global DMS emissions result in substantially lower $r_{\rm diff}$ values (${\sim}0\%$ and ${\sim}43\%$; Figures 4a-b) compared to a strictly dayside DMS emission assumption (${\sim}67\%$; Figure 4c). The larger value of $r_{\rm diff}$ in the latter simulation is due to the relatively short lifetime of DMS \citep{kloster2005dms}. 

Similar to the above DMS behavior, a potential consequence of tidal-locking is the relegation of CH$_4$ and N$_2$O production to a single hemisphere, i.e., processes of methanogenesis and denitrification favor anaerobic conditions and may be disfavored on photosynthetic oxygen-producing daysides. Spatially-variable surface to atmosphere flux distributions of CH$_4$ and N$_2$O therefore could exhibit higher $r_{\rm diff}$ than the values predicted here ($\la10\%$; Figures 1e-f and 1h-i).

\section{Discussion}

Here we discuss possible areas of future advancement, as well as the observational relevance of this study.

In this CCM study, we find that factors that determine biosignature concentration and distribution on a habitable tidally-locked planet are species dependent. For example, ozone mixing ratios are primarily driven by photolytic production and destruction, while ozone distribution and nightside sustenance are controlled by its transport and lifetime. An additional consideration, here demonstrated in our DMS simulations, is the spatial variance of gas fluxes. Given that habitable exoplanets are likely to possess heterogeneous ecologies, whose fluxes will interface with attendant atmospheric structure and circulation patterns, spatially-heterogeneous surface fluxes could have observationally-distinguishable effects on atmospheric spectra. For example, dayside upwelling could facilitate vertical mixing of surface gases into the upper atmosphere (Figure 1e-f), while nightside radiation inversions could trap constituents near the surface, limiting vertical mixing, day-nightside interactions, and potentially observability. These scenarios, in which atmospheric dynamics, photochemistry, surface flux sources, and feedback processes play important roles highlight the utility of 3D CCM simulations. However, due to non-linear interactions and internal atmospheric variability, disentangling drivers of emergent behavior is challenging and will likely require the tools of modern atmospheric and computational science, including Lagrangian tracking of constituents (e.g., \citealt{solch2010large}), single- and multi-model ensembles (e.g., \citealt{kay2015community}), and statistical analyses focused on detection and attribution (e.g., \citealt{horton2015contribution,diffenbaugh2017quantifying}).

Despite these challenges, the introduction of 3D CCMs to exoplanet biosignature prediction efforts offers substantial research potential. Future applications are likely to consider a wider variety of Earth-like biospheres, e.g., markedly disparate atmospheres of Earth throughout geologic time \citep{Rugheimer+Kaltenegger2018ApJ,ArneyEt2016AsBio}, and should be expanded to include biologically constrained models and modules \citep{CatlingEt2018AsBio,WalkerEt2018AsBio}. Such applications will facilitate egocentricity avoidance -- a commonly acknowledged goal of the field (e.g., \citealt{SeagerEt2013ApJ}). Until then, use of default Earth conditions may restrict the relevance of 3D CCM findings to truly Earth-like planets, with similar atmospheric formation histories, ecospheres, and biological signatures resulting from oxygenic photosynthesis \citep{MeadowsEt2016AsBio}. 

This study demonstrates that fully-coupled CCMs are particularly promising for studies that seek to assess the roles of and feedbacks between different stellar SEDs, biological behaviors, and atmospheric compositions. Such efforts are consistent with recent reviews, discussing aspirational goals and the future of exoplanet biosignature research \citep{CatlingEt2018AsBio,WalkerEt2018AsBio}. Extrasolar astrophysical radiation environments and/or atmospheric conditions may alter biological activity, as life is both photochemically and climatologically mediated. Living organisms are highly receptive toward UV emissions such that UV-B ($290< \lambda <320$ nm) photons hinder metabolism, photosynthesis, and thus biological production rates \citep{jansen1998higher}. Moreover, due to different climate and redox conditions, for example on anoxic Archean Earth (${\sim} 3.0$ Ga), hydrocarbons and organosulfur biosignatures (C$_2$H$_6$, CH$_4$, OCS, DMS etc.) could rise to more prominent abundances and hence may be conducive to remote detection \citep{Haqq-MisraEt2008AsBio,Domagal-GoldmanEt2011,HuEt2012ApJ,ArneyEt2016AsBio}. 

In terms of the potential observational implications of our simulations, our results agree with those of \citet{SeguraEt2005AsBio}, \citet{RauerEt2011A&A}, and \citet{RugheimerEt2015ApJ}, i.e., rocky planets orbiting active and quiet M-dwarfs should have deeper absorption depths, particularly for ozone and secondary biosignatures such as methane and nitrous oxide (as seen in transit and emission spectra). This makes habitable zone planets orbiting M-dwarfs favorable targets. 3D predictions from our CCM simulations may be confirmed by remote observations. Phase curve analysis can potentially resolve 3D atmospheric structures of super-Earth and Earth-sized terrestrial planets \citep{StevensonEt2014SCIENCE,Kreidberg+Loeb2016ApJL}. For example, thick substellar clouds could appear characteristically for planets with specific spit-orbit resonances \citep{YangEt2013ApJL}. 

In terms of biosignature measurements on tidally-locked planets, different longitudinal gradients of gaseous constituents may affect measurements of variation spectra (peak amplitude of the phase curves) extracted from thermal phase curves \citep{SelsisEt2011A&A}. As variation spectral signals depend on amplitude-peaks in orbital light curves, there may be added anisotropy due to time-varying longitudinal gas distributions on tidally-locked planets in each orbit (assuming null obliquity, as seen in Figure 1). Compared to non-tidally locked fast rotators (with similar stellar UV activity and orbital period), we predict that more pronounced absorption signals would be seen in variation spectrum on tidally-locked planets, driven by greater difference between maximum and minimum phase amplitudes due to uneven hemispheric gas distributions. Emission spectrum at maximum phase (direct line-of-sight) should correspondingly see similar behavior, at least for a few IR-windows (e.g., between 3-9 $\mu$m; \citealt{SelsisEt2011A&A}). For direct imaging, one possibility is that these features may be more prominent during certain orbital phases. Ozone observability, for example, may decrease during secondary eclipses as the dayside with reduced ozone abundance would be Earth-facing. Radiative transfer models, using our CCM results as inputs, will be needed to quantitatively assess observational prospects of the above.

\section{Conclusions}

This Letter reports numerical simulations using a coupled 3D CCM to explore global distribution of biosignature gases on Earth-like and tidally-locked planets as a function of stellar spectral type, stellar activity, and planetary rotation period. Qualitatively similar to 1D models, we find increased mixing ratios of biogenic compounds (e.g., O$_3$, CH$_4$, and N$_2$O) for both active and inactive M-dwarf SEDs. These increases are most pronounced for planets around quiet M-dwarfs. Even though the effects of tidal-locking are noticeable in our simulations, they are not yet discernable with current observational techniques, i.e., the primary biosignatures simulated in this work (O$_3$, CH$_4$, N$_2$O) show low ($\la 20\%$) day-to-nightside mixing ratio contrasts. Conversely, simulated day-to-nightside differences of photosynthetic compounds (e.g., DMS) are found to be nearly 70\% and underscore the need for heterogeneous 3D realism in modeling biosignatures and their photochemical derivatives. Overall, this study serves as a stepping stone for future applications using 3D CCMs to study the habitability and spectroscopic observability of terrestrial exoplanets. 

\acknowledgements
H.C. thanks J. Schnell and A. Loeb for helpful discussions. E.T.W. thanks NASA Habitable Worlds Grant 80NSSC17K0257 for support. Goddard affiliates are thankful for support from GSFC Sellers Exoplanet Environments Collaboration (SEEC), which is funded by the NASA Planetary Science Division’s Internal Scientist Funding Mode. The VPL team at University of Washington is thanked for the stellar spectra data.  Computational, storage, and staff resources were provided by the QUEST high performance computing facility at Northwestern University, which is jointly supported by the Office of the Provost, Office for Research, and Northwestern University Information Technology.

\end{document}